\documentstyle[prc,preprint,aps,epsf]{revtex}
\begin{document}
\draft
\title{Sum rules regarding the sign problem in Monte Carlo shell model 
calculations}

\author{C.~ W.~Johnson$^a$ and D.~J.~Dean$^b$}
 
\address{$^a$Department of Physics,
Louisiana State University, Baton Rouge Louisiana 70803\\
$^b$Physics Division, Oak Ridge National Laboratory, Oak Ridge, Tennessee
37831,\\
and Department of Physics and Astronomy, University of Tennessee, Knoxville,
Tennessee 37996
}

\maketitle

\begin{abstract}
The Monte Carlo shell model is a powerful technique for computational 
nuclear structure.  Only a certain class of nuclear interactions, however, 
such as pairing and quadrupole, are free 
of the numerical noise  known as the sign problem. 
This paper presents  sum rules that relate the sign problem to 
 the $J=0$ 
pairing matrix elements, thus illuminating the 
extrapolation procedure routinely used for realistic shell-model interactions.  
\end{abstract}


In recent years quantum Monte Carlo methods have become powerful numerical 
techniques for attacking the many-body problem in condensed matter, 
nuclear physics, and other fields.  Unfortunately, these methods are also  
suceptible to what is known as the sign problem.  Generically speaking, 
the sign problem arises when an expectation value to be computed is the average
of a wildly fluctuating quantity, so that 
numerical noise overwhelms the signal. 

This paper focuses on the auxiliary-field path integral, 
which is evaluated via Monte Carlo sampling, and further concentrates on the 
application to the nuclear shell model \cite{lang,extrapolate,koonin}.  
The nuclear shell model is an 
excellent venue for auxiliary-field Monte Carlo (AFMC) because one can 
prove that a large class of systems are 
free from the sign problem.   The `goodness' of an interaction 
with respect to the sign problem is related to the time-reversal 
properties of the effective one-body Hamiltonian \cite{lang}.

This  paper derives two sum rules for the time-reversal 
properties, and thus the sign-problem properties,
 of shell-model interactions. 
The sum rules  are  ultimately expressed in terms of the  
pairing matrix elements and prove two points of conventional, but 
previously unproved, wisdom in AFMC computations.  

\section{Auxiliary-field Path Integrals}
%
%
Traditional shell-model calculations diagonalize a Hamiltonian 
$\hat{H}$ in a many-body basis; 
the current best limit on model-space dimension is 
about $5 \times 10^8$ \cite{bigshell} although 
more typical spaces have basis dimensions of $10^5$-$10^7$. 
Rather than diagonalization, AFMC shell model calculations 
employ the imaginary-time evolution operator $\exp(-\beta \hat{H})$ 
to project out the ground state and evaluate thermally weighted 
expectation values. For the purposes of this paper is 
it sufficient to sketch the broad outlines of the method; for a 
detailed description the interested reader is referred to 
 \cite{lang,koonin}.

The nuclear  Hamiltonian $\hat{H}$ is usually two-body and 
can be written 
as quadratic in one-body operators 
\begin{equation}
\hat{H} = \sum_\alpha \epsilon_\alpha \hat{\rho}_\alpha
+ {1 \over 2} \sum_\alpha V_\alpha \hat{\rho}_\alpha^2.
\end{equation}
This is done in detail below. 
The quadratic term is source of all troubles; in order to 
facilitate numerical computation it is  linearized, 
via the Hubbard-Stratonovich 
transformation, to form an 
effective one-body Hamiltonian that depends on the imaginary time $\tau$:
\begin{equation}
\hat{h}(\tau) = \sum_\alpha (\epsilon_\alpha + s_\alpha V_\alpha 
\sigma_\alpha(\tau)) \hat{\rho}_\alpha,
\label{effham}
\end{equation}
where $s_\alpha = \pm 1$ if $V_\alpha < 0$ and $=\pm i$ if $V_\alpha > 0$. 
Then the imaginary-time evolution operator 
$\exp(-\beta \hat{H})$ becomes a path-integral:
\begin{equation}
\exp(-\beta \hat{H}) = 
\int {\cal D}[\sigma]
\exp \left( -{1\over 2} \int_0^\beta d\tau 
\sum_\alpha |V_\alpha|^2 \sigma_\alpha(\tau)^2 \right ) 
\left [ {\cal T} 
\exp \left ( - \int_0^\beta d\tau\hat{h}(\tau) \right) \right ],
\label{pathintegral}
\end{equation}
where ${\cal T}$ denotes time-ordering. This path integral 
can be evaluated by Monte Carlo sampling if a sufficiently well-behaved 
weight function $W(\sigma)$ can be found. An ideal weight function 
could be, for example, the trace of the integrand of Eqn.~(\ref{pathintegral})
 itself.  Unfortunately 
this choice of  $W(\sigma)$ often takes on both 
positive and negative values and one must use 
$|W|$ as a weight function. In those cases typically 
$\left \langle W/|W| \right \rangle \ll 1$, and the statistical 
errors for the Monte Carlo value of the observable are overwhelmingly 
large. For AFMC methods this is the manifestation of the sign problem.  

\section{Application to the spherical shell model}

The  shell model is built upon single-fermion states,  using
fermion creation and destruction operators $a^\dagger, a$.   
 For a  spherical basis 
these fermion states have good angular momentum $j$ and $m$, as well 
as orbital angular momentum $\ell$ which dictates the parity 
$(-1)^\ell$. 
For each state we associate a ``time-reversed'' partner, $j$ and $-m$, 
and define a time-reversal operation by a bar:   
$\bar{a}_{j,m} = (-1)^{j+m+\ell}a_{j,-m}$ and 
$\bar{a}^\dagger_{j,m} = (-1)^{j+m+\ell}a^\dagger_{j,-m}$.
Because of the half-spin statistics, 
$\bar{\bar{a}}_{j,m} = - a_{j,m}$. As hinted above, the 
time-reversal properties play a central role. 
 
The two-body  Hamiltonian is  usually written as 
\begin{equation}
\hat{H}_2 = {1 \over 4} \sum_{abcd}\zeta_{ab} \zeta_{cd} \sum_{JT} 
 \sum_{M T_z} V_{JT}(ab,cd) 
 \hat{A}^\dagger_{JT,MT_z}(ab)  \hat{A}_{JT,MT_z}(cd),
 \label{twobody}
\end{equation}
where $\zeta_{ab} \equiv \sqrt{1+\delta_{ab}}$, the $V_{JT}$ are 
matrix elements of the interaction, 
 and the two-body creation 
operator is 
$\hat{A}^\dagger_{JT,MT_z}(ab) = \left [ a^\dagger_{j_a} \otimes 
a^\dagger_{j_b} \right ]_{JT,MT_z}$.
Peforming a Pandya transformation from the particle-particle to the 
particle-hole representation, the Hamiltonian becomes 
\begin{equation}
\hat{H}_2 = {1 \over 2} \sum_{abcd} \sum_{KI} E_{KI}(ab,cd)
\sum_M (-1)^M \hat{\rho}_{KM,I}(ac) \hat{\rho}_{K-M,I}(bd) ,
\end{equation}
plus a residual one-body term that does not concern us here. 
The one-body or density operator is 
$\hat{\rho}_{KM}(ac) = \left [ a^\dagger_{j_a} \otimes 
\bar{a}_{j_c} \right ]_{KM}$.
and proton and neutron densities are combined:
\begin{equation}
\hat{\rho}_{KM,I} \equiv \hat{\rho}_{KM,p}+(-1)^I \hat{\rho}_{KM,n}.
\end{equation}
Note:  We will 
refer to these densities as `isoscalar' ($I=0$) and `isovector' ($I=1$)
even though the latter do not truly have good isospin.

The matrix elements in the particle-hole representation are 
\begin{eqnarray}
E_{K,I=0} (ac,bd) = (-1)^{j_b+j_c} \zeta_{ab} \zeta_{cd} \sum_J (-1)^J (2J+1) 
\left \{ 
\matrix{ j_a & j_b & J \cr
         j_d & j_c & K }
\right \} \nonumber \\
\times {1 \over 2} \left ( V^N_{J,T=1}(ab,cd) 
+  {1 \over 2} \left [  V^A_{J,T=0}(ab,cd) - V^S_{J,T=1}(ab,cd) 
\right ] \right ),  \\
E_{K,I=1} (ac,bd) = -(-1)^{j_b+j_c} \zeta_{ab} \zeta_{cd} \sum_J (-1)^J(2J+1) 
\left \{ 
\matrix{ j_a & j_b & J \cr
         j_d & j_c & K }
\right \} \nonumber \\
\times {1 \over 4}  \left [  V^A_{J,T=0}(ab,cd) - V^S_{J,T=1}(ab,cd) 
\right ] .
\end{eqnarray}
$V^N$ is a general two-body matrix element which can be separated  into 
 symmetric and antisymmetric parts, $V^N = V^S +V^A$; the symmetry 
properties are given by 
$V_{JT}^{S/A}(ab,cd)\equiv 
{1 \over 2} [ V^N_{JT}(ab,cd) \pm (-1)^{J+j_a+j_b +T-1} V^N_{JT}(ba,cd)]$.
Of course, only the antisymmetric matrix elements $V^A$ are physical 
and are used in Eqn. (\ref{twobody}); 
the symmetric matrix elements $V^S$, although unphysical, are a degree of 
freedom in the particle-hole representation and can be used 
to encourage convergence. 

The next step is to diagonalize the $E_{KI}$, which has eigenvalues 
$\lambda_{K,I}(\alpha)$ and eigenvectors $v_{KI,\alpha}(ac)$ . 
Define 
$\hat{\rho}_{KI}(\alpha) = \sum_{ac} v_{KI,\alpha}(ac) \hat{\rho}_{KI}
(ac)$;
then finally
\begin{equation}
\hat{H}_2 = {1 \over 2} \sum_{KI} \sum_{\alpha} \lambda_{KI}(\alpha)
\sum_M (-1)^M \hat{\rho}_{KM,I}(\alpha)  \hat{\rho}_{K-M,I}(\alpha) .
\label{diagonalform}
\end{equation}
Now the Hamiltonian is in the form of a sum of manifestly quadratic 
operators and the Hubbard-Stratonovich transformation can be applied.

\section{Lang's theorem and sign-problem-free interactions}

In some systems $W$ is positive-definite and there is no sign problem. 
 Lang's theorem \cite{lang} 
shows that for nuclear physics a wide class of systems are free from 
the sign problem.

Any one-body 
operator $\rho$ can be separated into time-even and time-odd 
parts: $\rho = {1\over 2}( \rho+\bar{\rho}) + {1\over 2}( \rho-\bar{\rho})$.
The basic statement of Lang's theorem is that, for even $N$, even $Z$ 
systems (or $N=Z$ odd), if the effective one-body Hamiltonian 
(\ref{effham}) is strictly time-even, then there is no sign problem; 
specifically, that 
 $\left  \langle \Psi_T\right | {\cal T} 
\exp \left ( - \int_0^\beta d\tau\hat{h}(\sigma) \right) 
\left  | \Psi_T \right \rangle \ge 0$. 
Since $i=\sqrt{-1}$ is also time-odd, this means that time-even 
operators $\rho_\alpha$ must have a real coefficient and time-odd 
operators an imaginary coefficient.  Because imaginary coefficients 
only enter if $V_\alpha > 0$, this means that time-even terms arise from 
attractive interactions and time-odd terms from repulsive interactions. 
One can easily show that 
$\bar{\rho}_{KM}(ac)=(-1)^{K+M} \pi(ac) \rho_{K,-M}(ac)$, 
where $\pi(ac)=$ parity of density operator $=(-1)^{\ell_a + \ell_c}$.
(The parity can only be negative when interactions cross major 
harmonic oscillator shells \cite{crossshell}. For a $0\hbar \omega$ 
model space, such as the $sd$ or $pf$ shells, the parity can be 
ignored.)  
Hence the criterion for a `good' interaction is 
\begin{equation}
(-1)^K \pi(\alpha)\lambda_{KI}(\alpha) < 0.
\label{good}
\end{equation}

In order to deal with realistic interactions  
\cite{crossshell,Wildenthal}  which modestly violate Lang's rule, Alhassid 
et al. \cite{extrapolate} introduced a method whereby one  
 deforms the realistic interaction to one that conforms to Lang's 
rule and then extrapolates from the ``good'' regime to the ``bad'' 
regime. 
That is, let 
\begin{equation}
\hat{H} = \hat{H}_{\rm good} + \hat{H}_{\rm bad}
\end{equation} 
where $\hat{H}_{\rm good}$ satisfies Lang's rule and 
$\hat{H}_{\rm bad}$ uniformly violates it.  This is done by 
decomposing the interaction into the form (\ref{diagonalform}), and 
segregrating `good' from `bad' interactions via (\ref{good}).
 Then one introduces 
\begin{equation}
\hat{H}(g) = f(g)\hat{H}_{\rm good} + g\hat{H}_{\rm bad},
\label{extrapolation}
\end{equation} 
where one usually takes $f(g)=[1 - (1-g)/4]$.
For $g=1$ we have the original Hamiltonian, but for $g \le 0$, 
$\hat{H}(g)$ satisfies Lang's rule and is amenable to the Monte Carlo 
shell model.  The typical procedure is to compute for several values 
of $g < 0$ and then extrapolate to $g = 1$.  
The results   agree for model spaces 
where exact diagonalizations can be performed for comparison 
\cite{extrapolate,koonin}. 
 
There are two points of conventional wisdom regarding Monte Carlo 
Shell Model calculations, both of which are noted in Lang et al 
\cite{lang} 
and later papers \cite{extrapolate,koonin}:

{\bf Conventional wisdom \# 1}:  It is customary to set the unphysical
$V^S_{J,T=1}$ to $ V^A_{J,T=0}$ so  
 that $E_{K,I=1} =0$. This reduces the number of 
auxiliary fields to be integrated over. On the other hand, it is 
conceivable that for some interactions one could exploit 
$E_{K,I=1} \neq 0$ to avoid the sign problem at the modest cost of 
doubling the number of auxiliary fields. 

This paper  proves explicitly that the latter statement is 
impossible and that the conventional choice is indeed the best, 
not only with regards to reducing the dimensions of the integral but 
with regard to alleviating the sign problem.

{\bf Conventional wisdom \# 2}.  The extrapolation procedure 
amounts to increasing the strength of the pairing interaction.  
Ref.~\cite{lang} notes that an attractive pairing interaction,
with $V^A_{J=0,T=1}(aa,cc) = - G \sqrt{(2j_a+1)(2j_c+1)}$
yields $\lambda_{K,I=0} = -(-1)^K G/2$ for all $K$; if added crudely 
in sufficient strength to any interaction it will make it `good.'
Empirically 
one finds that the more general 
extrapolation procedure described above seems to 
increase the pairing strength \cite{extrapolate,koonin}.  
The sum rules  given below show explicitly a relation between 
the sign problem and the  pairing interaction.

\section{Two sum rules and their implications}

This is the central section of the entire paper. 
The sum rule is, for isoscalar densities ($I=0$) is 
\begin{equation}
\sum_K (2K+1)\sum_\alpha (-1)^{K} \pi(\alpha)\lambda_{K,I=0}(\alpha) 
= 
 {1 \over 2} \sum_{a c } 
\pi(ac)\sqrt{ (2j_a+1) (2j_c+1) } V_{J=0,T=1}^A(aa,cc),
\end{equation}
and for isovector densities($I=1$)
\begin{equation}
\sum_K (2K+1)  \sum_\alpha (-1)^{K} \pi(\alpha) \lambda_{K,I=1}(\alpha) 
=0.
\end{equation}

The proof of these sum rules is straightforward.
 For a given $K$, $I$, 
$\sum_\alpha \lambda_{KI}(\alpha) = \sum_{ac} E_{KI}(ac,ac)$, 
since the sum of the eigenvalues is just the trace of the matrix. 
Then, using 
$$
\left \{ 
\matrix{ j_a & j_c & K \cr
         j_c & j_a & 0 }
\right \}
= {(-1)^{K+j_a + j_c}  \over \sqrt{ (2j_a+1) (2j_c +1) }},
$$
and the orthogonality of six-$j$ symbols \cite{edmonds}, one finds
\begin{eqnarray}
\sum_K (2K+1) (-1)^{K} \sum_\alpha\lambda_{K,I=0}(\alpha) 
=\nonumber  \\ {1 \over 2} \sum_{ac}  
\left ( V^N_{J=0,T=1}(aa,cc) 
+  {1 \over 2} \left [  V^A_{J=0,T=0}(aa,cc) - V^S_{J=0,T=1}(aa,cc) 
\right ] \right ), \label{sum1} \\
\sum_K (2K+1) (-1)^{K} \sum_\alpha \lambda_{K,I=1}(\alpha) 
=-{1 \over 4} \sum_{ac} 
 \left [  V^A_{J=0,T=0}(aa,cc) - V^S_{J=0,T=1}(aa,cc) 
\right ] . \label{sum2}
\end{eqnarray}
(For each of these steps 
parity must be conserved so that implicitly $\pi(\alpha) = \pi(ac)$ in 
equations (\ref{sum1}), (\ref{sum2}). ) 
The final step is to note that, due to their symmetry properties, 
 $V_{J=0,T}(aa,cc)$ must be 
symmetric for $T=0$ and antisymmetric for $T=1$.  Hence the terms 
$V^A_{J=0,T=0}(aa,cc)$, $V^S_{J=0,T=1}(aa,cc)$ vanish and 
$V^N_{J=0,T=1}(aa,cc)$ reduces to $V^A_{J=0,T=1}(aa,cc)$. Thus 
the final results follow easily.

Now to interpret these sum rules. 
Recall that Lang's rule dictates that a `good' interaction will 
have $(-1)^{K} \pi(\alpha) \lambda_{KI} (\alpha) \leq 0$.  By inspecting 
the `isovector' 
sum rule, if any $\lambda_{K,I=1}$ is `good' (and nonzero) 
there must be another, nonzero `bad' $\lambda$ to cancel 
its contribution.  The only option to 
avoid the sign problem in the isovector channel is to set all 
$\lambda_{K,I=1}=0$, thus confirming conventional wisdom \# 1.
 
The isoscalar sum rule is  expressed in terms of the $J=0$ 
pairing matrix elements.  
First consider `normal parity' pairing matrix elements, that is, 
within a major oscillator shell. 
If $H_{bad} \rightarrow -H_{bad}$, 
then the contribution of the `bad' $\lambda$'s will switch 
signs and increase the magnitude of the isoscalar sum rule. 
This must then manifest as stronger, more attractive
 pairing matrix elements, as expounded in conventional wisdom \#2. 
The `abnormal parity' pairing matrix elements, which takes
a $J=0$ pair of nucleons from one oscillator shell to one of 
opposite parity, must however become more repulsive. 

Such behavior is illustrated in Figure 1 for a 
 realistic cross-shell interaction \cite{crossshell}.
There we plot the $J=0$ matrix elements $V_{J=0 T=1}(ab,cd)$, 
grouped by shells, denoting the $sd$-shell by $+$    and $pf$-shell
by $-$.
The case $g=1$ (filled circles) 
correspond to matrix elements from the original, realistic 
interaction that has a `bad' sign, while $g=0$ (open circles) 
correspond to an interaction minimally deformed to have a `good' sign.
As described by the isoscalar sum rule, the intershell 
($V(++,++)$ and $V(--,--)$ )  pairing 
becomes more attractive as $g =1 \rightarrow 0$, while the intrashell 
($V(++,--)$) becomes more repulsive.  Matrix elements 
of the type $V(+-,+-)$ do not exhibit any significant trend, 
which is accord with the sum rules:  such matrix elements do not 
contribute to the trace over the $E_K(ac,ac)$.

In summary, the two sum rules derived herein allow one to confirm the 
conventional wisdom regarding AFMC computations.  They also potentially 
yield  a powerful tool to analyze the extrapolation and to 
look for alternatives.  This is left to future work.

This work was done with the support of the Department of Energy under 
grants 
DE-FG02-96ER40985 and DE-FG02-97ER40963.   
Oak Ridge National Laboratory is managed by Lockheed Martin Energy
Research Corp. for the U.S. Department of Energy under contract number
DE-AC05-96OR22464.

\bibliographystyle{try}

\begin{figure}
\caption{ 
$J=0$ pairing matrix elements for realistic interaction 
in $sd$-$pf$ shell model calculations. $g=1$ corresponds to  
original, realistic Hamiltonian that has a `bad' sign; $g=0$ is 
an interaction minimally deformed to have a `good' sign. $x$-axis 
is arbitrary ordering of $J=0$ matrix elements. `$+$' denotes the 
$sd$ shell while `-' denotes the $pf$ shell.
}

 \begin{center}
${\epsfxsize=140mm\epsffile{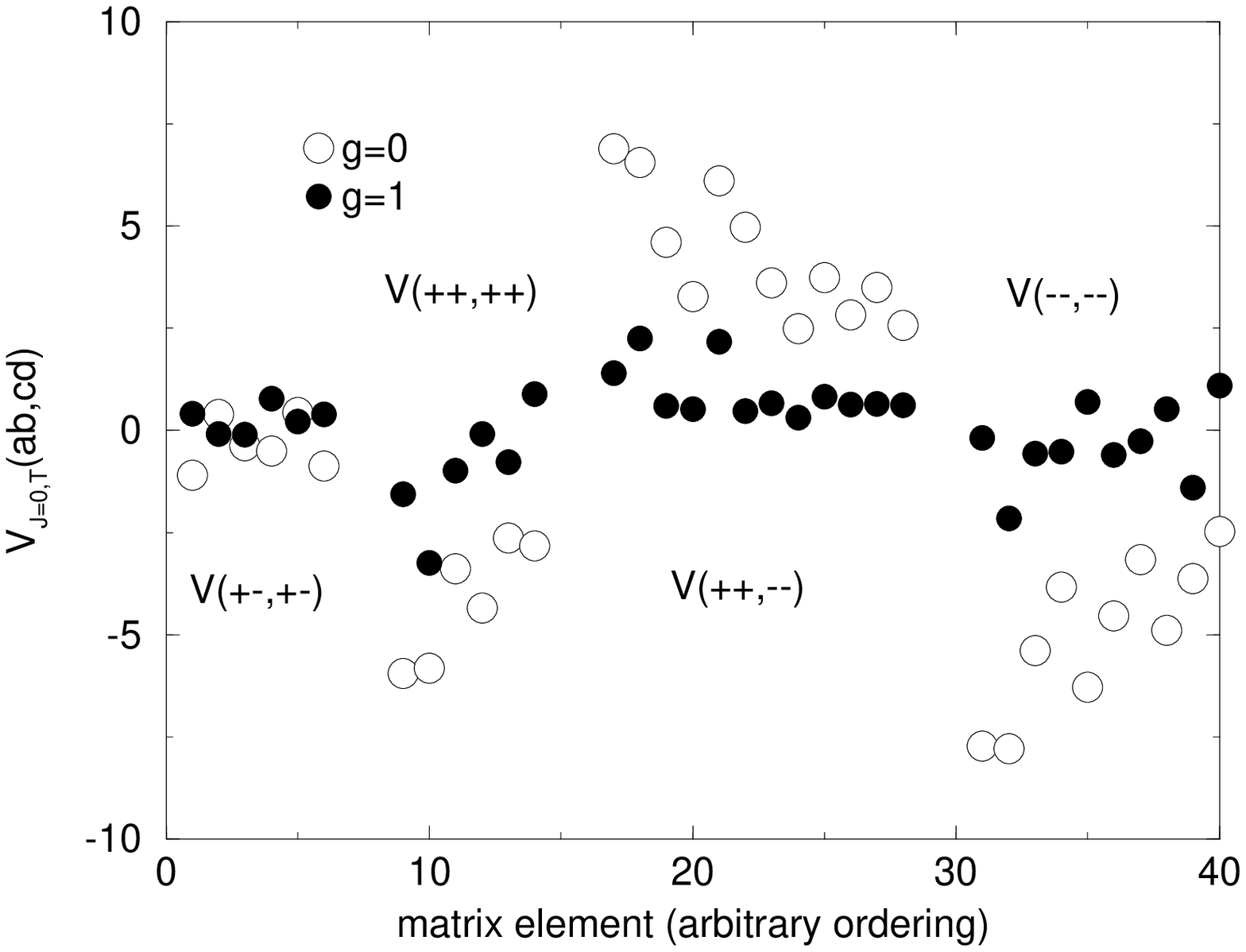}}$

    \end{center}
\end{figure}

\end{document}